\documentclass[10pt]{iopart}

\usepackage{amssymb}
\usepackage{graphicx}

\newcommand{\bC} {\mathbb{C}}
\newcommand{\bN} {\mathbb{N}}
\newcommand{\bR} {\mathbb{R}}
\newcommand{\cH} {\mathcal{H}}
\newcommand{\cP} {\mathcal{P}}

\newcommand{\fexp}  [1] {\exp \left( #1 \right)}
\newcommand{\fabs}  [1] {\left| #1 \right|}
\newcommand{\fsqrt} [1] {\sqrt{#1} \,}
\newcommand{\fnorm} [1] {\left\| #1 \right\|}
\newcommand{\fnormq}[1] {{\left\| #1 \right\|}^2}

\newcommand{\eqref}[1]{(\ref{#1})}

\newcommand{\mc}{\, mc}
\newcommand{\mcq}{\, mc^2}
\newcommand{\Ang}{\,\mbox{\AA}}

\begin{document}


\title{Relativistic Time-of-Arrival and Traversal Time}
\author{Andreas Ruschhaupt}
\address{Faculty of Physics, University of Bielefeld,
Universit\"atsstr. 25,\\
D-33615 Bielefeld, Germany}
\ead{rushha@physik.uni-bielefeld.de}

\begin{abstract}
We compute in a relativistic way the time-of-arrival and the traversal time
through a region
of a free particle with spin $\frac{1}{2}$. We do this by applying
the relativistic extension of the Event-Enhanced Quantum Theory which we
have presented in a previous paper. We find a very good coincidence of
the results of our formalism and the results obtained by using
classical relativistic mechanics.
\end{abstract}



\section{Introduction}
\label{sec1}

Blanchard and Jadczyk \cite{blanchard.1993a,
blanchard.1995c,blanchard.1995d} have proposed an extension of standard
(non-relativistic) quantum mechanics called Event-Enhanced Quantum
Theory (EEQT) which main idea is to view the total system as
consisting of coupled classical and quantum part. The pure states
of the quantum part are wave functions which
are not directly observable, whereas the pure states of the classical part can 
be observed without disturbing them. Changes of the classical pure
states are discrete and irreversible, they are called events.
A review about applications of EEQT is for example
\cite{blanchard.2000}.

Blanchard and Jadczyk have also introduced a relativistic
extension of EEQT \cite{blanchard.1996b} using the idea of a proper time
and an indefinite scalar product.

In a previous paper \cite{ruschhaupt.2002a}, we have presented an
alternative approach for a relativistic extension of EEQT. We
summarize this approach in \Sref{sec2}.

The question when a particle arrives at a given point cannot be
answered unambiguously in the standard 
formulation of quantum mechanics. Nevertheless, there exists a lot of approaches to
answer the question of ``time-of-arrival''. A review about
``time-of-arrival'' can be for example found in Muga, Sala, and Palao
\cite{muga.1998} and an extensive review including a lot of references
is Muga and Leavens \cite{muga.2000}.

In this paper, we consider a two dimensional spacetime. We examine
the following operational definition of the time-of-arrival:
The particle is prepared in a space-time point
$(0,x_0)$ and moves freely (except of the influence exerted on it by the detector)
in positive $x$ direction. A detector is put at $x_D$ with
$x_D > x_0$. It measures the time-of-arrival of the particle at $x_D$.
Because it is possible that the particle is never detected, the
experiment or simulation should be stopped after a reasonable and finite period of time.

This definition of time-of-arrival has been examined in the framework
of non-relativistic EEQT \cite{blanchard.1996a}. In \Sref{sec4} of this paper, we
will compute the time-of-arrival using our relativistic extension of EEQT
and we will compare the results to those obtained by using classical
relativistic mechanics of a point particle.

The question how long a particle needs to traverse a given finite region
cannot also be
answered unambiguously in the standard 
formulation of quantum mechanics. This time is called traversal time.
It is often examined if a potential is in the given region and the particle must
tunnel through the region. Reviews about ``traversal time'' and ``tunnelling time''
including a lot of references are for example \cite{hauge.1989,
landauer.1994,nimtz.1997, chiao.1997}. 

``Tunnelling time'' has been also examined using the non-relativistic
EEQT \cite{palao.1997, ruschhaupt.1998, ruschhaupt.2000}.

In \Sref{sec5} of this paper, we examine the ``free traversal time'' for an
always freely moving particle (again except of the influence exerted on the particle
by the detectors).

Using two detectors at rest one behind the other, we can measure the traversal
time through the region located between the two detectors:
The particle is again prepared in a space-time point $(0,x_0)$ and moves in
positive $x$-direction.
We put a detector $D_1$ at $x_1$ with $x_1 > x_0$. This detector can detect the
particle without destroying it. A second detector $D_2$ is put at $x_2$
with $x_2 > x_1$. It destroys the particle after detection.

At the beginning of the measurement, both detectors $D_1$ and $D_2$
are active. If detector $D_1$ detects the particle (without destroying it), it turns
itself off, but detector $D_2$ stays turned on. If detector $D_2$ detects
the particle, the experiment is finished. 

Thus, the particle can be detected by detector $D_1$ at a time $t_1$ and then by
detector $D_2$ at a time $t_2$. If this happens, the time difference $t_2 - t_1$
is defined to be the ``traversal time''.

It is also possible that the particle is detected by $D_2$ without prior detection by
$D_1$, but this situation should not contribute to traversal times.

Moreover, it is possible that the particle is never detected or only
detected one time by detector
$D_1$. For this reason the experiment or simulation should be stopped after a
reasonable and finite period of time.

We will compute the traversal time using
our relativistic extension of EEQT. Again we will compare the results to those
obtained using classical relativistic mechanics of a point particle.


\section{A Relativistic Extension of EEQT}
\label{sec2}

We recall the extension of EEQT which we have proposed in
\cite{ruschhaupt.2002a}. It describes
one spin $\frac{1}{2}$-particle in a relativistic way and in four
dimensional spacetime. Here, we restrict ourself to consider two
dimensional spacetime.

As in EEQT, we postulate that the total system consists of a classical
and a quantum part which are coupled. Because of that, at a given
proper time $\tau$, the (pure) state of
the total system is a pair $(\omega_\tau,\Psi_\tau)$. $\omega_\tau$ 
is the state of the classical part and $\Psi_\tau$ is the state of the
quantum part.

We assume that a (pure) state  $\omega_\tau$ of the classical part is a number:
$\omega_\tau \in \bN_0 = \{0,1,2,...\}$. We also call a change of the
classical (pure) state ``event'' as in non-relativistic EEQT.

The (pure) states of the quantum part shall be (heuristically) solutions
$\Psi : \bR \times \bR \to \bC^4$ of the Dirac equation. Because we
examine only free particles in this paper, we use the free Dirac equation:
\begin{eqnarray}
\rmi \hbar c \frac{\partial}{\partial(ct)} \Psi (ct,x) = H_0 \Psi
(ct,x)
\label{sec2_dirac}
\end{eqnarray}
with $H_0 = -\rmi c \hbar \gamma^0 \gamma^1 \frac{\partial}{\partial
x} + mc^2 \gamma^0$. We denote the space of quantum states by $\cH$.
A more precise definition of $\cH$ can be found in \cite{ruschhaupt.2002a}.
We use the Dirac or standard representation of the $\gamma$-matrices.
Let $\cP  =  \left\{ (y^0,y^1,\alpha) : y^0,y^1,\alpha \in \bR , 
\fabs{\alpha} < 1 \right\}$ and 
$\sigma_\lambda (u) = \left( y^0 + \alpha \cdot u, \; y^1 + u \right)$, 
$\forall u \in \bR, \forall \lambda \equiv (y^0,y^1,\alpha)~\in~\cP$.
We now introduce a positive-definite scalar product between two quantum states
$\Psi_A,\Psi_B \in \cH$:
\begin{eqnarray}
<\Psi_A | \Psi_B>_{\cH} \; := \int_{\sigma_\lambda} j^\mu_{AB} \rmd f_\mu
\end{eqnarray}
with $\lambda \in \cP$, $j^\mu_{AB} = \Psi_A^+ \gamma^0 \gamma^\mu \Psi_B$ and $df_\mu =
(1,-\alpha) du$ denotes the differential ``surface'' element of
$\sigma_\lambda$.
This scalar product is well defined because it is independent of
$\lambda$. This follows from Gauss theorem and the fact that
$\partial_\mu j^\mu_{AB} = 0$. Moreover, one can show that this
scalar product is covariant, its value being independent of the reference frame.

We introduce the operators $U_{(ct_0, x_0)}$ with $ct_0,x_0 \in \bR$:
\begin{eqnarray*}
(U_{(ct_0, x_0)}\Psi) (x) := \Psi (ct_0, x_0 + x)
\end{eqnarray*}
An interesting property of a quantum state is that it is uniquely given by its
values on a space-like hyperplane $\sigma_\lambda$.
Therefore, the operators $U_{(ct_0, x_0)}$ are invertible.
$\Psi = U^{-1}_{(ct_0,x_0)} \psi$ is the solution of the
free Dirac equation \eqref{sec2_dirac} fulfilling the
initial condition $\Psi (ct_0, x) = \psi (x - x_0)$. We
get
\begin{eqnarray*}
\Psi (ct,x) = (U^{-1}_{(ct_0,x_0)} \psi) (ct,x) =
\fexp{-\frac{\rmi}{\hbar}(t-t_0) H_0} \psi(x - x_0)
\end{eqnarray*}

Now, we want to formulate an algorithm for modelling continuous relativistic
measurements, indeed we
will propose in the following an algorithm to simulate detections of the particle.
In principle, we will do this by rewriting the algorithm of EEQT,
replacing $t$ with $\tau$ and using our Hilbert space of
``solutions.''

We denote the reference frame by $K$. The particle is prepared at proper
time $\tau_0$ in a point $(ct_0,x_0)$.

We consider $n$ detectors with trajectories $z_j (\tau)$, $j=1..n$. The
trajectories start at proper time $\tau = \tau_0$ from the backward light-cone of the
space-time point of the ``preparation event'':
\begin{eqnarray*}
\left(ct_0 - z_j^0(\tau_0)\right)^2 - \left( x_0 - z_j^1(\tau_0)\right)^2 = 0, 
\quad z_j^0(\tau_0) \le ct_0
\end{eqnarray*}
 
We allow detections which happen in the past of the preparation time.
But we do not allow detections, if the detection space-time point is located in the
backward light-cone of the space-time point of the preparation event.

Each detector is characterized by operators $G_j (\tau) : \cH \to \cH$. Let
$G_j^+ (\tau)$ be the adjoint operator. The total coupling between the quantum
and the classical part is given by $\Lambda(\tau) := \sum_{j=1}^n G_j^+ (\tau) G_j (\tau)$.

We define a detection algorithm in the following way:
\begin{itemize}
\item[(i)] The particle is prepared in a space-time point $(ct_0,x_0)$ at proper time
  $\tau=\tau_0$. The quantum state is $\Psi_{\tau_0}$ with
  $\fnormq{\Psi_{\tau_0}}_{\cH} \equiv
  \, <\Psi_{\tau_0}|\Psi_{\tau_0}>_{\cH} = 1$ and
  the classical state is $\omega_{\tau_0} = 0$.

\item[(ii)] Choose uniformly a random number $r \in [0,1]$.

\item[(iii)] Propagate the quantum state forward in proper time by solving
\begin{eqnarray}
  \frac{\partial}{\partial \tau} \Psi_\tau = -\frac{1}{2} \Lambda(\tau)
  \Psi_\tau
\label{sec2_dgl}
\end{eqnarray}
until $\tau = \tau_1$, where $\tau_1$ is defined by
\begin{eqnarray*}
  1 - \fnormq{\Psi_{\tau_1}}_{\cH} \, = \int_{\tau_0}^{\tau_1} \rmd\tau
  <\Psi_\tau|\Lambda\Psi_\tau>_{\cH} \, = r
\end{eqnarray*}
Let $\omega_\tau = \omega_{\tau_0}$ until $\tau=\tau_1$, a
detection happens at proper time $\tau = \tau_1$.

\item[(iv)] We choose the detector $k$ - which detects the particle - with
  probability
\begin{eqnarray*}
  p_k = \frac{1}{N} \fnormq{G_k(\tau_1) \Psi_{\tau_1}}_{\cH}
\end{eqnarray*}
with $N = \sum_{j=1}^n \fnormq{G_j (\tau_1)\Psi_{\tau_1}}_{\cH}$.

\item[(v)] Let $l$ be the detector which detects effectively the particle. The
  detection happens at the point $z_l (\tau_1)$. The detection induces the
  following change of the states:
\begin{eqnarray*}
 \left(\omega_{\tau_1}, \Psi_{\tau_1} \right) & \longrightarrow & 
 \left(l, \frac{G_l (\tau_1) \Psi_{\tau_1}}
  {\fnorm{G_l (\tau_1) \Psi_{\tau_1}}_{\cH}} \right)
\end{eqnarray*}
\end{itemize}
The algorithm can start again perhaps with other detectors at position (ii).

Because the scalar product is covariant, this algorithm is
covariant. Moreover, its non-relativistic limit reduces to the algorithm of the
non-relativistic EEQT. If we ``charge conjugate'' the initial state
$\Psi_{\tau_0} \to \Psi_{\tau_0}^C \equiv C {\gamma^0}^T \Psi_{\tau_0}^*$ and the detector
functions $G_j(\tau) \to C {\gamma^0}^T G^*_j (\tau) {\gamma^0}^T
C^+$ with $C=\rmi \gamma^2\gamma^0$, then the algorithm will give
the same detections as if we start with $\Psi_{\tau_0}$ and
$G_j(\tau)$ (if we choose the same random numbers).
The quantum state in the ``charge conjugated'' world $\Psi^C_\tau$ and the
quantum state in the ``normal'' world are always connected by $\Psi^C_\tau
= C {\gamma^0}^T \Psi^*_\tau$.

Note, that we have also formulated an algorithm for modelling ideal measurements
of infinitesimal small duration in \cite{ruschhaupt.2002a}. It can be
seen as playing the role of a relativistic, covariant reduction postulate.


\section{Initial quantum state}
\label{sec3}

The particle is prepared at proper time $\tau = \tau_0$ in the space-time
point $(0,x_0)$ with a mean momentum $p_0$.
We examine three different initial states of the particle in this paper. 
Remember that a initial states of the particle must be a solution of
the Dirac equation \eqref{sec2_dirac}.

The first state corresponds to a state with only positive energies:
\begin{eqnarray*}
\fl \Psi_{0,P} (ct,x) =
\frac{1}{N_P} \int \rmd k \,
\frac{1}{2\hat{E}} \cdot F_{\Delta k} \left(k - \frac{p_0}{\hbar} \right)
\left(\begin{array}{c}\hat{E}+\hat{m}\\0\\0\\k\end{array}\right)
\cdot \fexp{\rmi k (x-x_0) - \rmi \hat{E} ct}
\end{eqnarray*}
with $\hat{m} = \frac{mc}{\hbar}$, $\hat{E} =
\fsqrt{k^2+\hat{m}^2}$, $\Delta k = 10 \Ang^{-1}$,
\begin{eqnarray*}
F_{\Delta k} (k) = \left\{ \begin{array}{cl} \fexp{-\frac{k^2}{\Delta k^2 - k^2}} 
& \mbox{for} \; \fabs{k} < \Delta k \\
0 & \mbox{otherwise} \end{array} \right.
\end{eqnarray*}
and $N_P$ being a normalization factor so that
$\fnormq{\Psi_{0,P}}_{\cH} = 1$. This state describes an electron with
charge $-e$.

The second one corresponds to a state with only negative energies:
\begin{eqnarray*}
\fl \Psi_{0,N} (ct,x) =
\frac{1}{N_N} \int \rmd k \,
\frac{1}{2\hat{E}} \cdot F_{\Delta k} \left(k - \frac{p_0}{\hbar} \right)
\left(\begin{array}{c}\hat{E}-\hat{m}\\0\\0\\k\end{array}\right)
\cdot \fexp{-\rmi k (x-x_0) + \rmi \hat{E} ct}
\end{eqnarray*}
with $N_N$ being a normalization factor so that
$\fnormq{\Psi_{0,N}}_{\cH} = 1$.
Remember that the above algorithm is invariant under charge
conjugation. If we consider the charge conjugate of the initial state and the
detector functions, we get the same events. Because the ``charge
conjugated'' world and the ``normal world'' should describe the same
physical situation and because the charge conjugation of $\Psi_{0,N}$
describes a particle with charge $+e$ in the ``charge conjugated
world'', we demand that the initial state $\Psi_{0,N}$ describes a
positron with charge $+e$ also in the ``normal'' world.

As third initial state, we want to use a mixed state:
\begin{eqnarray*}
\fl \Psi_{0,PN} (ct,x)
= U^{-1}_{(0, x_0)} \left[ \frac{1}{(2\pi)^{1/4}\fsqrt{\eta}} \cdot
    \fexp{-\frac{x^2}{4\eta^2} + \rmi \frac{p_0}{\hbar}x}\cdot
    \left(\begin{array}{c}1\\0\\0\\0\end{array}\right) \right] (ct,x)\\
\fl = \frac{\fsqrt{2\eta}}{(2\pi)^{3/4}} \int \rmd k \,
\frac{1}{2\hat{E}} \fexp{-\eta^2 \left(k - \frac{p_0}{\hbar} \right)^2}
\left(\begin{array}{c}\hat{E}+\hat{m}\\0\\0\\k\end{array}\right)
\cdot \fexp{\rmi k (x-x_0) - \rmi \hat{E} ct}\\
\fl + \frac{\fsqrt{2\eta}}{(2\pi)^{3/4}} \int \rmd k \,
\frac{1}{2\hat{E}} \fexp{-\eta^2 \left(k + \frac{p_0}{\hbar} \right)^2}
\left(\begin{array}{c}\hat{E}-\hat{m}\\0\\0\\k\end{array}\right)
\cdot \fexp{-\rmi k (x-x_0) + \rmi \hat{E} ct}
\end{eqnarray*}
with $\eta = 0.1 \Ang$. The constants are chosen in such a way that
$\fnormq{\Psi_{0,PN}}_{\cH} = 1$.
The physical interpretation of the mixed state is the following:
we assume that the particle (a single particle) can be in an ``electron-state''(solution with
positive energies) and in a
``positron-state'' (solution with negative energies), in analogy to the case, that a
particle can be e.g. in a spin 
$+\frac{1}{2}$-state or in a spin $-\frac{1}{2}$-state. Superpositions
as $\Psi_{0,PN}$ of the two states should be (in analogy to the spin-case)
possible and allowed.


\section{Free Time-of-Arrival}
\label{sec4}

In this section, we apply the above algorithm to simulate the detection
of the particle by one detector which is at rest.

We want to compare the results to those which we obtain by using
classical relativistic mechanics of a point-particle. 

Let us use the reference frame $K_0$ in which the detector is at
rest. In this reference frame, the particle is prepared at proper time
$\tau_0=0$ in the space-time point $(0,x_0)$ with a mean momentum $p_0$.
The detector is put at $x_D$, its trajectory is
$z(\tau) = (c\tau + x_0 - x_D, x_D)$.
The coupling operator should be given by
\begin{eqnarray*}
  G (\tau) = U^{-1}_{z (\tau)} g(x) U_{z(\tau)}
\end{eqnarray*}
with $g(x)$ characterizing the sensitivity of the detector:
\begin{eqnarray*}
  g (x) = \fsqrt{\frac{2W_D}{\hbar}} \cdot
  F_{\frac{\Delta x_D}{2}} (x)
\end{eqnarray*}
The adjoint operator is $G^+ (\tau) = U^{-1}_{z(\tau)} g^+(x)
 U_{z(\tau)}$.

Because it is possible that the particle is never detected, we stop the
algorithm at $\tau=\tau_{CUT}$ (with $\tau_{CUT}$ large).

We want to recall that the algorithm is covariant. The choice of $K_0$ as the reference
frame is arbitrary. The algorithm can be applied in any reference frame, and
it will result (if we choose the same random numbers) the same events in all
reference frames.

Using our algorithm, the probability that the detector detects the
particle at all is given by
\begin{eqnarray*}
  P_{\infty} = \int_0^{\tau_{CUT}} \rmd\tau \, <\Psi_\tau | \Lambda \Psi_\tau >_{\cH}
\end{eqnarray*}
The probability density for a ``proper time-of-arrival'' at the detector is
given by ($\tau < \tau_{CUT}$)
\begin{eqnarray*}
  p(\tau) = \frac{1}{P_\infty} <\Psi_\tau | \Lambda \Psi_\tau >_{\cH}
\end{eqnarray*}
It is zero for $\tau \le 0$ and $\tau \ge \tau_{CUT}$.

Using this probability density for ``proper time-of-arrival'', we can calculate
the probability density and the 
expectation value for the time-of-arrival in an arbitrary reference frame.

Let us first look at the detector's rest-frame $K_0$. If a detection happens at
proper time $\tau$, then it happens in space-time point $z(\tau) = (c \tau+x_0-x_D,
x_D)$. This implies a time-of-arrival of $t = \tau - \frac{x_D-x_0}{c}$. So we get
the following probability density for the time-of-arrival in the
detector's rest-frame $K_0$:
\begin{eqnarray*}
  \varrho_0 (t) & = & p \left( t + \frac{x_D - x_0}{c} \right)
\end{eqnarray*}
The expectation value (or mean time-of-arrival) is 
\begin{eqnarray*}
\fl T_{a,0} & = & \int \rmd t \, t \, \varrho_0 (t) = \int \rmd\tau
  \, \left( \tau - \frac{x_D-x_0}{c} \right) p(\tau) = \int \rmd\tau \, \tau \, p(\tau) -
  \frac{x_D-x_0}{c}
\end{eqnarray*}

Now, we want to calculate these values in a reference frame $K_v$
which moves with velocity $v$ with respect to the detector's rest-frame $K_0$.
The Poincar\'e-transformation $K_0 \to K_v$ has the following form:
\begin{eqnarray*}
  \tilde{x} = \frac{1}{\fsqrt{1-\frac{v^2}{c^2}}} \left(
\begin{array}{cc} 1&-\frac{v}{c}\\-\frac{v}{c}&1\end{array}
\right) x
\end{eqnarray*}
The detector trajectory in $K_v$ is
\begin{eqnarray*}
\fl \tilde{z} (\tau) =
\left(1-\frac{v^2}{c^2} \right)^{-\frac{1}{2}} \cdot (c\tau + x_0 -
x_D - v/c \cdot x_D \; , \quad -v\tau - v/c \cdot x_0 + v/c \cdot x_D + x_D)
\end{eqnarray*}
So the normalized probability density for the time-of-arrival in the reference 
frame $K_v$ is given by
\begin{eqnarray*}
  \varrho_v (\tilde{t}) & = & \fsqrt{1-\frac{v^2}{c^2}} \cdot p
  \left( \fsqrt{1-\frac{v^2}{c^2}} \tilde{t} + \frac{x_D-x_0}{c} +
  \frac{v}{c^2} x_D \right)
\end{eqnarray*}
and the expectation value (or mean time-of-arrival) in $K_v$ is given by
\begin{eqnarray}
  T_{a,v} & = & \int \rmd\tilde{t} \, \tilde{t} \, \varrho_v (\tilde{t}) 
= \frac{1}{\fsqrt{1-\frac{v^2}{c^2}}} \left[ T_{a,0} -
\frac{v}{c^2}x_D \right]
\label{sec4_conn_diff_RF}
\end{eqnarray}


\subsection{Numerical Approach}
\label{sec4_num}

We use the reference frame $K_0$ to compute $p(\tau)$. 
Therefore, we define
\begin{eqnarray*}
  \Omega (\tau, x) := (U_{z(\tau)} \Psi_\tau) (x) = \Psi_\tau (c \tau +
  x_0 - x_D, x_D + x)
\end{eqnarray*}
If $\Psi_\tau$ is a solution of \eqref{sec2_dirac} and
\eqref{sec2_dgl}, then we get
\begin{eqnarray}
\fl \rmi \hbar \frac{\partial}{\partial \tau}\Omega (\tau,x)\nonumber\\
\fl =  \rmi \hbar c \frac{\partial}{\partial (ct)} \Psi_\tau (\overbrace{c\tau
 + x_0 - x_D}^{ct}, x_D + x) + \rmi \hbar \left( \frac{\partial \Psi_\tau}{\partial
 \tau} \right) (c\tau + x_0 - x_D , x_D + x) \nonumber \\
\fl = 
  H_0 \Omega (\tau,x) - \rmi
  \frac{\hbar}{2} g^+(x) g(x) \Omega(\tau,x)
\label{sec4_dgl_omega}
\end{eqnarray}
We have to solve this equation with the initial condition 
$\Omega (0,x) = \Psi_0 (x_0-x_D, x_D+x)$.
Using $<\Psi_\tau|\Lambda \Psi_\tau>_{\cH} \, = \int \rmd x \, \Omega^+(\tau,x) g^+(x)
g(x)\Omega(\tau,x)$, we can calculate $P_\infty$ and $p(\tau)$ if we
know $\Omega (\tau,x)$. Using $p(\tau)$, we get $\varrho_0(t)$ and $T_{a,0}$.

The equation \eqref{sec4_dgl_omega} with the initial condition $\Omega (0, x) = 
 \Psi_0 (x_0 - x_D, x_D + x)$ is solved numerically. The proper time dynamics
 of $\Omega$ is approximated by 
\begin{eqnarray*}
\fl \Omega(\tau+\Delta \tau) \approx \fexp{-\frac{\Delta\tau}{2} \frac{1}{2} g^+(x) g(x)}
  \fexp{-\Delta\tau \frac{\rmi}{\hbar} H_0} 
  \fexp{-\frac{\Delta\tau}{2}\frac{1}{2} g^+(x) g(x)}\Omega (\tau)
\end{eqnarray*}
We now discretize the proper time and the space with steps 
$\Delta x_B = c \Delta \tau_B = 0.0004 \Ang$. Then, the first and the last
operator can be computed directly. The second operator is discretized by using
the method of Wessels, Caspers, and Wiegel \cite{wessels.1999}. The
boundary conditions are walls at $x=-6 \Ang$ and 
at $x=4 \Ang$ in such a way that $\Omega (\tau, -6\Ang) = \Omega (\tau, 4 \Ang)
= 0$ for all $\tau$. We set: 
$\tau_{CUT} = 13.0\Ang/c \, (p_0<0.5 \mc)$, 
$\tau_{CUT} = 7.0 \Ang/c \, (0.5\mc \le p_0 < 0.75\mc)$,
$\tau_{CUT} = 5.0 \Ang/c \, (0.75\mc \le p_0 < 1.0\mc)$,
$\tau_{CUT} = 4.5 \Ang/c \, (1.0\mc \le p_0)$.
We do the simulations again with other time and space steps
$\Delta x_A = c \Delta \tau_A = 0.0006 \Ang$. So the error in the expectation
value $T_{a,0}$ can be approximated by
\begin{eqnarray}
error(T_{a,0}) = \pm \frac{\Delta x_B}{\Delta x_A - \Delta x_B} \fabs{T_{a,0}(\Delta x_B) -
    T_{a,0}(\Delta x_A)}
\label{sec4_error_T}
\end{eqnarray}


\subsection{Results}

We set $x_0 = -1 \Ang$, $x_D=0 \Ang$, $\Delta x_D=0.01\Ang$ and
$W_D = 1 \times 10^{-5} \mcq$.

\begin{figure}[t]
  \begin{flushright}
  \includegraphics [width=0.8\linewidth]{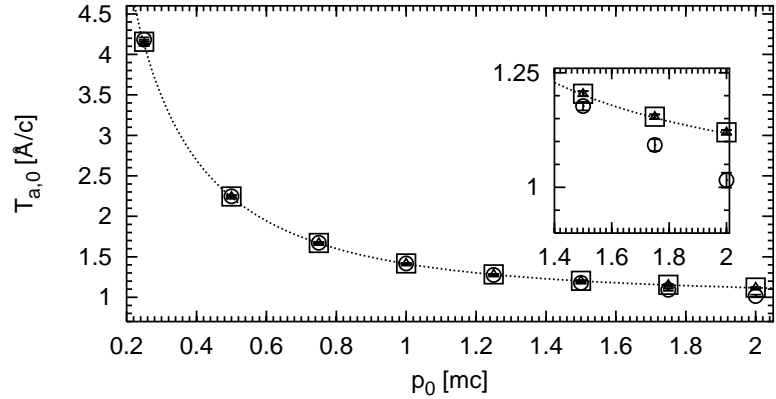}
  \end{flushright}
  \caption{\label{fig_1}Mean time-of-arrival $T_{a,0}$ versus particle momentum $p_0$ in the
      detector's rest-frame $K_0$, relativistic
      simulation with detector parameters $\Delta x_D= 0.01 \Ang$, $W_D=1\times
      10^{-5} \mcq$
      started with different initial states : 
      $\Psi_{0,P}$ (boxes with error bars), 
      $\Psi_{0,N}$ (triangles with error bars), 
      $\Psi_{0,PN}$ (circles with error bars), other
      parameters see text; classical relativistic mechanics $t_{a,RM}$ (dotted line);
      the figure inside is a zoom of the right lower area of the figure outside}    
\end{figure}

\Fref{fig_1} shows the corresponding expectation values of the
time-of-arrival $T_{a,0}$ in the detector's rest-frame $K_0$ for different momenta
$p_0$ and for the three different initial states. The error bars are
calculated using \eqref{sec4_error_T}.

In addition, \Fref{fig_1} shows the arrival-times calculated by using the
classical relativistic mechanics of a point-particle:
\begin{eqnarray*}
t_{a,RM} = \frac{x_D - x_0}{c} \fsqrt{1 + \frac{m^2 c^2}{p_0^2}}
\end{eqnarray*}

We see that the expectation values are nearly independent of the initial state $\Psi_{0,PN}$,
$\Psi_{0,P}$ or $\Psi_{0,N}$. Furthermore, it exists a good agreement between the
values we computed and the results obtained by using classical relativistic mechanics.

Only for very high momenta, the expectation values of
the simulation with $\Psi_{0,PN}$ are a bit smaller than the times from
classical mechanics and those obtained by the simulations with other initial states.

\begin{figure}[t]
  \begin{flushright}
    \includegraphics [width=0.75\linewidth]{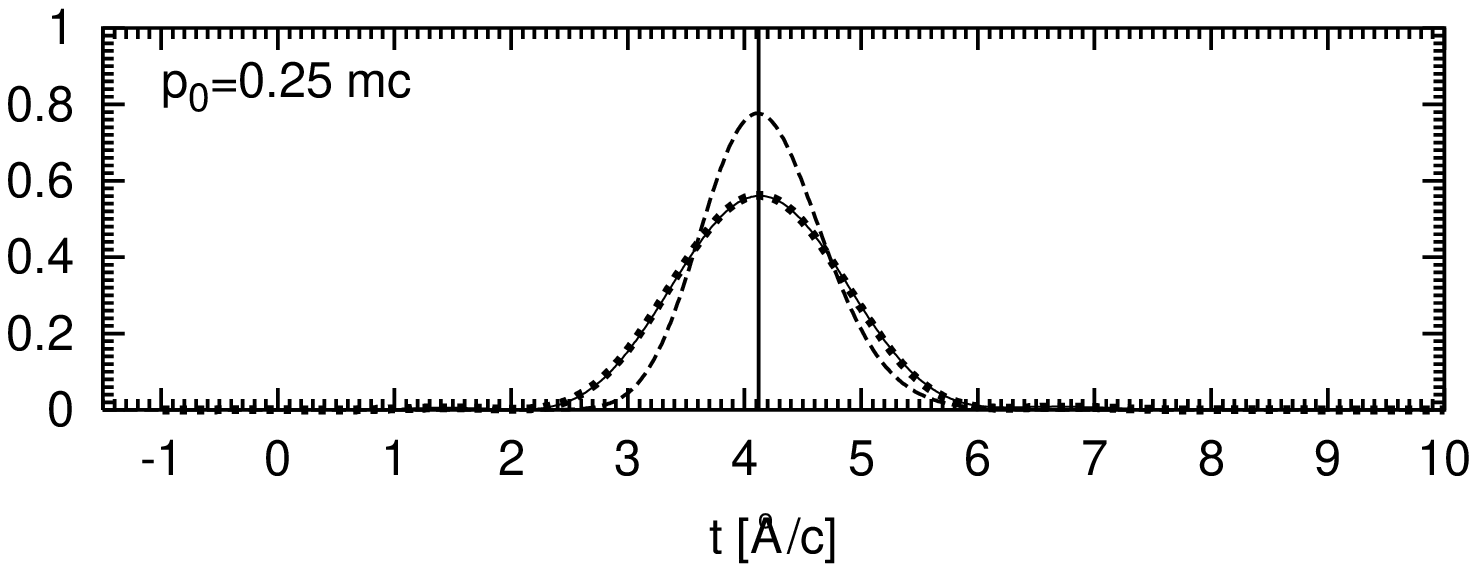}\\
    \includegraphics [width=0.75\linewidth]{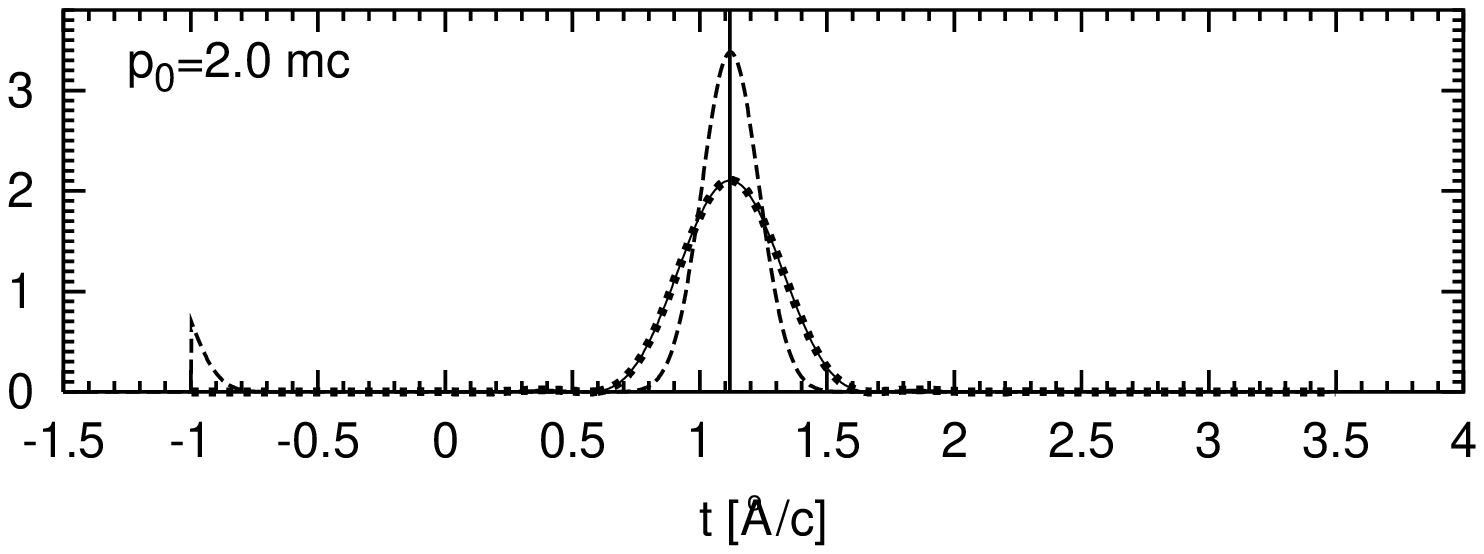}
  \end{flushright}
  \caption{\label{fig_2}Probability densities  $\varrho_0$ for the time-of-arrival in the
      detector's rest-frame $K_0$, 
      detector parameters: $\Delta x_D = 0.01 \Ang$, $W_D=1\times
      10^{-5} \mcq$,
      initial state:  
      $\Psi_{0,P}$ (small solid line), 
      $\Psi_{0,N}$ (big dotted line),
      $\Psi_{0,PN}$ (dashed line),
      particle momentum $p_0$; the
      vertical solid line indicates the arrival time deduced from classical relativistic
      mechanics}
\end{figure}

The reason can be seen in \Fref{fig_2}, which shows probability densities
in the detector's rest-frame $K_0$. For very high momenta and if we
start with $\Psi_{0,PN}$, we find a small probability for negative
times-of-arrival. This fact explains why the expectation values of the
simulation with $\Psi_{0,PN}$ are smaller than the results of classical
mechanics and those of the other simulations.

We also see that the probability densities are (nearly) the same if we start with
$\Psi_{0,P}$ or $\Psi_{0,N}$.

The expectation values $T_{a,v}$ in different reference frames are connected by
\eqref{sec4_conn_diff_RF}.

We get from classical relativistic mechanics: the time-of-arrival $\tilde{t}_{a,RM}$
in the reference frame $K_v$ is connected to the result $t_{a,RM}$ in the
reference frame $K_0$ in the same manner:
\begin{eqnarray*}
  \tilde{t}_{a,RM} = \frac{1}{\fsqrt{1-\frac{v^2}{c^2}}} 
\left[ t_{a,RM} - \frac{v}{c^2} x_D \right]
\end{eqnarray*}
  
So we have a good agreement between the simulated expectation values and
the results deduced from classical relativistic mechanics in all reference
frames!

Another important and interesting question is how the expectation
values depend on the parameters of the
detector. The initial state is now the function $\Psi_{0,P}$ with positive
energies. We examine those particle momenta which are also examined in
\Fref{fig_1}. We compute the probability densities and the
expectation values for four different pairs of detector parameters.

First, we examine the case of a ``higher'' detector with $W_D=1.0 \mcq$,
but its width is still $\Delta x_D = 0.01 \Ang$. We find that the
expectation values and the normalized probability densities do not change by using
a ``higher'' detector for all examined particle momenta. The detection
probability $P_\infty$ increases with increasing detector height $W_D$.

Next, we examine the case of a wider detector with $\Delta x_D= 0.4 \Ang$ and
$W_D=1\times 10^{-5}\mcq$. The expectation values do not
change. The normalized probability density do not change in a significant way, it only
becomes a bit wider. 
Again the detection probability $P_\infty$ increases with increasing
detector width $\Delta x_D$.

The results only change if we use a very wide and height detector with $\Delta
x_D= 0.4 \Ang$ and $W_D= 1.0 \mcq$. The expectation values and the normalized
probability densities are then shifted to earlier times.

In other words, the simulations show that it exists a wide range of detector
parameters for which the results do not change significantly.


\section{Free Traversal Time}
\label{sec5}

In this section, we simulate the traversal time measurement described
in \Sref{sec1} by applying our algorithm being described in
\Sref{sec2}.

We use the detectors' rest-frame $K_0$.
The particle is prepared at proper time $\tau_0 = 0$ in $(0, x_0)$. The first
detector $D_1$ is at rest at position $x_1$. Its
trajectory is $z_1 (\tau) = (c\tau + x_0 - x_1, x_1)$. The second
detector $D_2$ is at rest at position $x_2$. Its
trajectory is $z_2 (\tau) = (c\tau + x_0 - x_2, x_2)$.
The coupling operators of detector $D_j$ should be given by
\begin{eqnarray*}
  G_j (\tau) = U^{-1}_{z_j (\tau)} g_j(x) U_{z_j(\tau)} \qquad j=1,2
\end{eqnarray*}
with $g(x)$ characterizing the sensitivity of the detector $D_j$:
\begin{eqnarray*}
  g_j (x) = \fsqrt{\frac{2W_j}{\hbar}} \cdot
  F_{\frac{\Delta x_j}{2}} (x)
\end{eqnarray*}

Because it is possible that the particle is not detected two times, we stop the
algorithm at $\tau=\tau_{CUT}$ (with $\tau_{CUT}$ large).

Let $\Psi_0$ be the initial
state and $\Psi_\tau$ the solution of \eqref{sec2_dirac} and
\eqref{sec2_dgl}. Then, 
the probability that the particle is detected by $D_1$ at all is
\begin{eqnarray*}
P_{\infty,1} = \int_0^{\tau_{CUT}} \rmd\tau \, 
<\Psi_\tau| G_1^+ (\tau) G_1 (\tau) \Psi_\tau>_{\cH}
\end{eqnarray*}
The probability 
density that the particle is detected by $D_1$ is given by ($\tau < \tau_{CUT}$) 
\begin{eqnarray*}
p_1 (\tau) = \frac{1}{P_{\infty,1}} <\Psi_\tau| G_1^+ (\tau) G_1 (\tau) \Psi_\tau>_{\cH}
\end{eqnarray*}
If a detection by detector $D_1$ happens at $\tau_1$, the quantum state
after the detection is given by:
\begin{eqnarray}
\Phi_{\tau_1}^{(\tau_1)} := \frac{G_1 (\tau_1) \Psi_{\tau_1}}
  {\fnorm{G_1 (\tau_1) \Psi_{\tau_1}}_{\cH}}
\label{sec5_tra_3}
\end{eqnarray}
Let $\Phi_\tau^{(\tau_1)}$ be the solution of \eqref{sec2_dgl} with initial state
\eqref{sec5_tra_3}. 
We get the following conditional probability that the particle is detected a second
time by $D_2$ if it is detected by $D_1$ at $\tau_1$:
\begin{eqnarray*}
P_\infty^{(\tau_1)} = \int_{\tau_1}^{\tau_{CUT}} \rmd\tau_2 \,
<\Phi_{\tau_2}^{(\tau_1)}| G_2^+ (\tau_2) G_2 (\tau_2) \Phi_{\tau_2}^{(\tau_1)}>_{\cH}
\end{eqnarray*}
The probability density for a second detection at proper time $\tau_2$
by detector $D_2$ after a detection of detector $D_1$ at proper time $\tau_1$ is
given by
\begin{eqnarray*}
p_2^{(\tau_1)} (\tau_2) = \frac{1}{P_\infty^{(\tau_1)}}
<\Phi_{\tau_2}^{(\tau_1)}| G_2^+ (\tau_2) G_2 (\tau_2) \Phi_{\tau_2}^{(\tau_1)}>_{\cH}
\end{eqnarray*}
Finally, the probability density for a first detection by $D_1$ at $\tau_1$ and a
second detection by $D_2$ at $\tau_2$ is
\begin{eqnarray*}
\fl p_{12} (\tau_1, \tau_2) = 
\frac{p_2^{(\tau_1)} (\tau_2) \cdot P_\infty^{(\tau_1)} \cdot p_1(\tau_1)
\cdot P_{\infty,1}}
{\int_0^{\tau_{CUT}} \rmd\tau \, P_\infty^{(\tau)} \cdot p_1 (\tau)
\cdot P_{\infty,1}} \\
\fl = \frac{1}{P_{\infty,12}} 
\cdot \left\{\begin{array}{ll}
\lefteqn{<\Phi_{\tau_2}^{(\tau_1)}| G_2^+ (\tau_2) G_2 (\tau_2) \Phi_{\tau_2}^{(\tau_1)}>_{\cH}
 \cdot <\Psi_{\tau_1}| G_1^+ (\tau_1) G_1 (\tau_1)
 \Psi_{\tau_1}>_{\cH}} & \\ 
& \mbox{for} \; 0 < \tau_1 < \tau_{CUT}\; \mbox{and} \; \tau_1 < \tau_2 <
\tau_{CUT} \hspace{4cm}\\
\\
0 & \mbox{otherwise}
\end{array} \right.
\end{eqnarray*}
with $P_{\infty,12}$ being the probability that the particle is detected
two times:
\begin{eqnarray*}
\fl P_{\infty,12} = P_{\infty,1} \cdot \int_0^{\tau_{CUT}} \rmd\tau \, P_\infty^{(\tau)} \cdot 
p_1 (\tau) \\ 
\fl = \int_0^{\tau_{CUT}} \!\!\!\! \rmd\tau_1 
\int_{\tau_1}^{\tau_{CUT}} \!\!\!\! \rmd\tau_2 \, 
<\Phi_{\tau_2}^{(\tau_1)}| G_2^+ (\tau_2) G_2 (\tau_2) \Phi_{\tau_2}^{(\tau_1)}>_{\cH}
 \cdot <\Psi_{\tau_1}| G_1^+ (\tau_1) G_1 (\tau_1) \Psi_{\tau_1}>_{\cH}
\end{eqnarray*}

Note, that this probability density is independent of the reference frame in
which the algorithm is applied.

We now calculate traversal times in different reference frames. In contrast to
$p_{12}$, the probability density for traversal time depends on the reference
frame.

We start with the detectors' rest-frame $K_0$. If the first detection of $D_1$ happens at
proper time $\tau_1$, then it happens at space-time point 
$z_1(\tau_1) =(c\tau_1 + x_0 - x_1, x_1)$. If the second detection of $D_2$ happens
at proper time $\tau_2$, then it happens at space-time 
$z_2 (\tau_2) = (c\tau_2 + x_0 - x_2, x_2)$. The resulting traversal
time is therefore
\begin{eqnarray*} 
t= \tau_2 + \frac{x_0 - x_2}{c} - \tau_1 - \frac{x_0 - x_1}{c} = \tau_2 -
\tau_1 - \frac{x_2 - x_1}{c}
\end{eqnarray*}
So the normalized probability density for the traversal time in the
detectors' rest-frame $K_0$ is given by
\begin{eqnarray*}
\rho_0 (t) = \int \rmd\tau \;
p_{12} \left(\tau, t + \frac{x_2-x_1}{c} + \tau \right)
\end{eqnarray*}
The expectation value of the traversal time (or mean traversal time)
in $K_0$ is
\begin{eqnarray*}
T_{t,0} = \int \rmd t \, t \cdot \int \rmd\tau \; p_{12} \left(\tau, t + \frac{x_2-x_1}{c}
+ \tau \right)
\end{eqnarray*}

Now, we want to calculate these values in the reference frame $K_v$ (the reference
frame which moves with velocity $v$ with respect to the detectors' rest-frame
$K_0$).

The detector trajectories in $K_v$ are
\begin{eqnarray*}
\tilde{z_1} (\tau) = \frac{1}{\fsqrt{1 - \frac{v^2}{c^2}}} \cdot \left(c\tau + x_0 -
x_1 - \frac{v}{c} x_1, \, -v\tau - \frac{v}{c}(x_0-x_1) + x_1 \right) \\
\tilde{z_2} (\tau) = \frac{1}{\fsqrt{1 - \frac{v^2}{c^2}}} \cdot \left(c\tau + x_0 -
x_2 - \frac{v}{c} x_2, \, -v\tau - \frac{v}{c}(x_0-x_2) + x_2 \right)
\end{eqnarray*}
If the first detection of $D_1$ happens at $\tau_1$ and the second detection of
$D_2$ happens at $\tau_2$, then it results a traversal time of
\begin{eqnarray*}
\tilde{t} = \frac{1}{\fsqrt{1 - \frac{v^2}{c^2}}} \cdot \left(\tau_2 - \tau_1 -
\frac{x_2 - x_1}{c} - \frac{v}{c^2} (x_2 - x_1) \right)
\end{eqnarray*}
So the normalized probability density for the traversal time in the
reference frame $K_v$ is given by
\begin{eqnarray*}
\fl \rho_v (\tilde{t}) = \fsqrt{1 - \frac{v^2}{c^2}} \cdot \int \rmd\tau \;
p_{12} \left(\tau, \fsqrt{1 - \frac{v^2}{c^2}} \tilde{t} + 
\frac{x_2-x_1}{c} + \frac{v}{c^2} (x_2 - x_1) + \tau \right)
\end{eqnarray*}
The expectation value of the traversal time (or mean traversal time) in $K_v$ is
\begin{eqnarray}
\fl T_{t,v} = \int \rmd t \, t \cdot \fsqrt{1 - \frac{v^2}{c^2}} \cdot \int \rmd\tau \;
p_{12} \left(\tau, \fsqrt{1 - \frac{v^2}{c^2}} \tilde{t} + 
\frac{x_2-x_1}{c} + \frac{v}{c^2} (x_2 - x_1) + \tau \right) \nonumber \\
\fl = \frac{1}{\fsqrt{1-\frac{v^2}{c^2}}} \left[ T_{t,0} - 
  \frac{v}{c^2} (x_2 -x_1) \right]
\label{sec5_Tv_T0}
\end{eqnarray}


\subsection{Numerical Approach}

We use the reference frame $K_0$.
For computation of the algorithm until the first detection, we define
\begin{eqnarray*}
\Omega_A (\tau,x) := (U_{(c\tau+x_0-x_1,0)} \Psi_\tau) (x) = \Psi_\tau (c\tau+x_0 -
x_1, x)
\end{eqnarray*}
$\Psi_\tau$ should be a solution of \eqref{sec2_dirac} and \eqref{sec2_dgl}, so we get
\begin{eqnarray}
\fl \rmi \hbar \frac{\partial}{\partial \tau}\Omega_A (\tau,x)
 = H_0 \Omega_A 
- \rmi \frac{\hbar}{2} g_1^+(x-x_1) g_1(x-x_1) \Omega_A \nonumber\\
 - \rmi \frac{\hbar}{2}
 T g_2(x-x_2)^+ g_2(x-x_2) T^{-1} \Omega_A
\label{sec5_dgl_omega_A}
\end{eqnarray}
with
$T= U_{(c\tau+x_0-x_1,0)} U^{-1}_{(c\tau+x_0-x_2,0)}=\fexp{-(x_2-x_1)\frac{\rmi}{c\hbar}H_0}$.
A solution of this equation has to be found satisfying the initial condition
$\Omega_A (0,x)=\Psi_0 (x_0-x_1, x)$.
The equation \eqref{sec5_dgl_omega_A} is solved numerical with the proper time
dynamics approximated by 
\begin{eqnarray*}
\fl \Omega_A(\tau+\Delta \tau)
 \approx\\
\fl \fexp{- \frac{\Delta\tau}{2} \frac{\imath}{\hbar} mc^2 \gamma^0
   - \frac{\Delta\tau}{2} \frac{1}{2} g_1^+(x-x_1) g_1(x-x_1)}
  \fexp{-\frac{\Delta\tau}{2}
    \frac{\imath}{\hbar}\left(-\imath\hbar c \gamma^0 \gamma^1
    \frac{\partial}{\partial x}\right)}\\
\fl T \fexp{-\Delta\tau\frac{1}{2} g_2^+(x-x_2) g_2(x-x_2)} T^{-1}
 \fexp{-\frac{\Delta\tau}{2}
    \frac{\imath}{\hbar}\left(-\imath\hbar c \gamma^0 \gamma^1
    \frac{\partial}{\partial x}\right)}\\
\fl \fexp{- \frac{\Delta\tau}{2} \frac{\imath}{\hbar} mc^2 \gamma^0
  - \frac{\Delta\tau}{2} \frac{1}{2} g_1^+(x-x_1) g_1(x-x_1)}  \Omega_A (\tau)
\end{eqnarray*}
with $T \approx \prod \fexp{- \Delta\tau \frac{\imath}{\hbar} H_0}$.

We discretize the proper time and the space with steps
$\Delta \tau$ and $\Delta x$ 
($c\Delta\tau=\Delta x$). The boundary conditions are walls at
$x=-8\Ang$ and $x=8\Ang$ in such a way
that $\Omega_A (\tau, -8\Ang) = \Omega_A (\tau, 8 \Ang) = 0$ for all $\tau$.

All operators (including $T$) can be evaluated directly or are
approximated by using the method of Wessels, Caspers, and Wiegel
\cite{wessels.1999} or by using Wendroff's formula (see e.g. \cite{mitchell.book}).

For simulating the second part of the algorithm (after a first
detection by detector $D_1$ at proper time $\tau_1$), we define
\begin{eqnarray*}
\Omega_B^{(\tau_1)} (\tau,x) := (U_{(c\tau+x_0-x_2, 0)} \Psi_\tau) (x) 
= \Psi_\tau (c\tau+x_0 - x_2, x)
\end{eqnarray*}
with $\Psi_\tau$ being a solution of \eqref{sec2_dirac} and \eqref{sec2_dgl}.
We get
\begin{eqnarray}
\fl \rmi \hbar \frac{\partial}{\partial \tau}\Omega_B^{(\tau_1)} (\tau,x)
 = H_0 \Omega_B^{(\tau_1)}(\tau,x) - 
  \rmi \frac{\hbar}{2} g_2^+(x-x_2) g_2(x-x_2) \Omega_B^{(\tau_1)} (\tau,x)
\label{sec5_dgl_omega_B}
\end{eqnarray}
We must solve this equation with the initial condition
\begin{eqnarray}
\fl \Omega_B^{(\tau_1)} (\tau_1,x) = \frac{T^{-1} g_1 (x-x_1) \Omega_A(\tau_1,x)}
{\fsqrt{\int \rmd x \, \Omega_A^+(\tau_1,x) g_1^+(x-x_1) g_1(x-x_1) \Omega_A(\tau_1,x)}}
\label{sec5_init_omega_B}
\end{eqnarray}
The equation \eqref{sec5_dgl_omega_B} with the initial condition
\eqref{sec5_init_omega_B} can be solved approximately in analogy to \Sref{sec4_num}.

Knowing $\Omega_A (\tau,x)$ and $\Omega_B^{(\tau_1)}(\tau,x)$, we can
calculate $P_{\infty,12}$ and $p_{12} (\tau_1, \tau_2)$. So we can
calculate $\rho_0(t)$ and $T_{t,0}$.

We do the computation with proper time and space step $c\Delta \tau_B = \Delta
x_B = 0.0006$. The value of $\tau_{CUT}$ depends on the particle
momentum: 
$\tau_{CUT}=31.5 \Ang/c \, (p_0=0.25 \mc)$, 
$\tau_{CUT}=17.5 \Ang/c \, (p_0=0.5 \mc)$,
$\tau_{CUT}=13.5 \Ang/c \, (p_0=0.75 \mc)$, 
$\tau_{CUT}=11.5 \Ang/c \, (1.0 \mc \le p_0 < 1.5 \mc)$,
$\tau_{CUT}=10.5 \Ang/c \, (1.5 \mc \le p_0)$.

Moreover, we do the computation with proper time and space step $c\Delta\tau_A =
\Delta x_A = 0.001$. So the error in the expectation value $T_{t,0}$
can be approximated by
\begin{eqnarray}
error(T_{t,0}) = \pm\frac{\Delta x_B}{\Delta x_A - \Delta x_B} \fabs{T_{t,0}(\Delta x_B) -
    T_{t,0}(\Delta x_A)}
\label{sec5_error_T}
\end{eqnarray}

The error in the probability $P_{\infty,12}$ is approximated by a similar
formula:
\begin{eqnarray}
error(P_{\infty,12}) = \pm\frac{\Delta x_B}{\Delta x_A - \Delta x_B}
    \fabs{P_{\infty,12}(\Delta x_B) - P_{\infty,12}(\Delta x_A)}
\label{sec5_error_P}
\end{eqnarray}


\subsection{Results}

We perform the simulation with different initial states and different particle
momenta $p_0$. We set $x_0 = -1.5 \Ang$.
The detector parameters are $x_1 = 0 \Ang$, $\Delta x_1 = 0.5 \Ang$, $W_1 = 1 \times
10^{-3} mc^2$ and $x_2 = 1.26 \Ang$, $\Delta x_2 = 0.02 \Ang$, $W_2 = 1 \times 10^{-3}
mc^2$.

\begin{figure}[t]
  \begin{flushright}
  \includegraphics [width=0.8\linewidth]{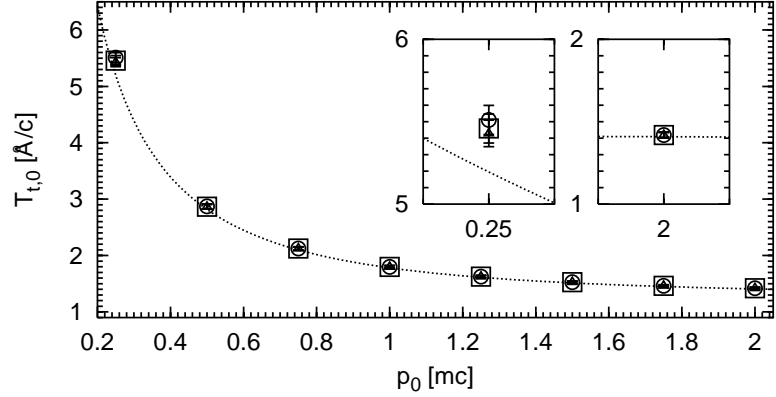}
  \end{flushright}
  \caption{\label{fig_3}Mean traversal time $T_{t,0}$ versus particle momentum $p_0$ in the
      detectors' rest-frame $K_0$, starting with different initial states : 
      $\Psi_{0,P}$ (boxes with error bars), 
      $\Psi_{0,N}$ (triangles with error bars), 
      $\Psi_{0,PN}$ (circles with error bars), 
      other parameters see text; results from classical relativistic
      mechanics $t_{t,RM}$ (dotted line)}
\end{figure}

Now, we are interested in the traversal time in the detectors' rest-frame $K_0$. 
\Fref{fig_3} shows the expectation values for traversal time with
different initial states and different particle momenta $p_0$. The errors
calculated by \eqref{sec5_error_T} are also plotted. 

A first result is that we see nearly no dependence on the initial state.

In addition, the times which one obtains by using classical relativistic
mechanics of a point-particle are plotted:
\begin{eqnarray*}
t_{t,RM} = \frac{x_2-x_1}{c} \cdot \fsqrt{1+\frac{m^2c^2}{p_0^2}}
\end{eqnarray*}

There is a good agreement between the simulated results and those
obtained by using classical relativistic mechanics. This agreement becomes
more accurate with increasing particle momentum $p_0$.

\begin{figure}[t]
  \begin{flushright}
    \includegraphics [width=0.75\linewidth]{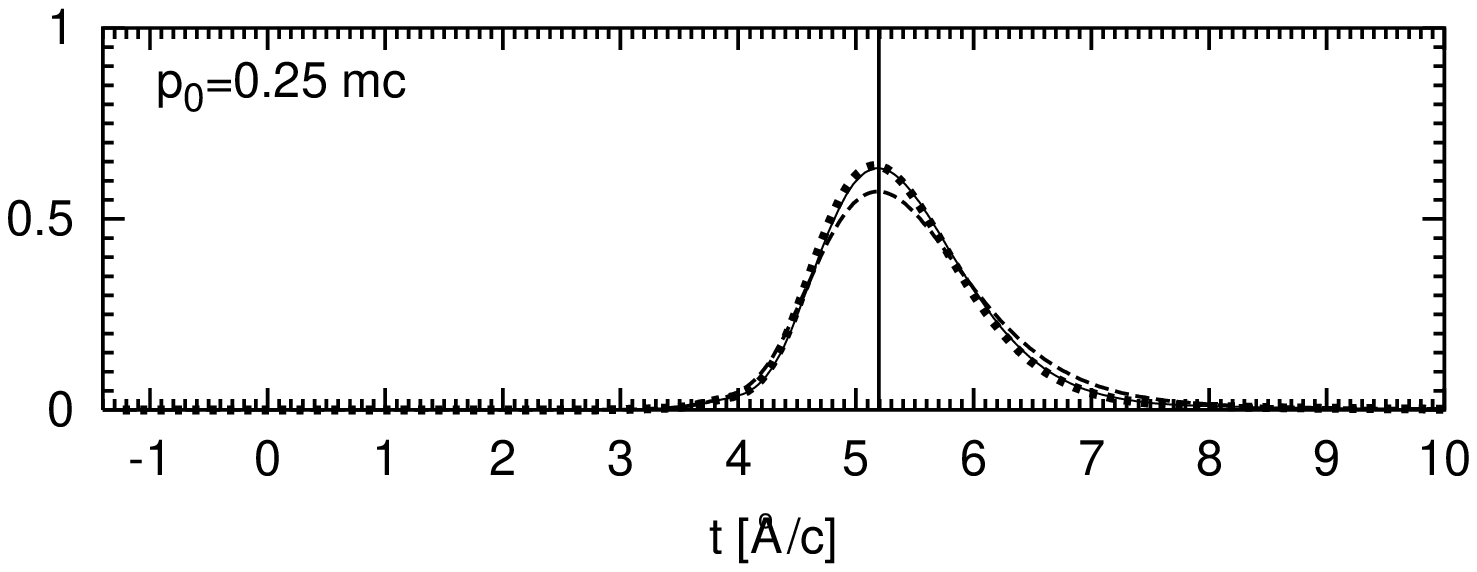}\\
    \includegraphics [width=0.75\linewidth]{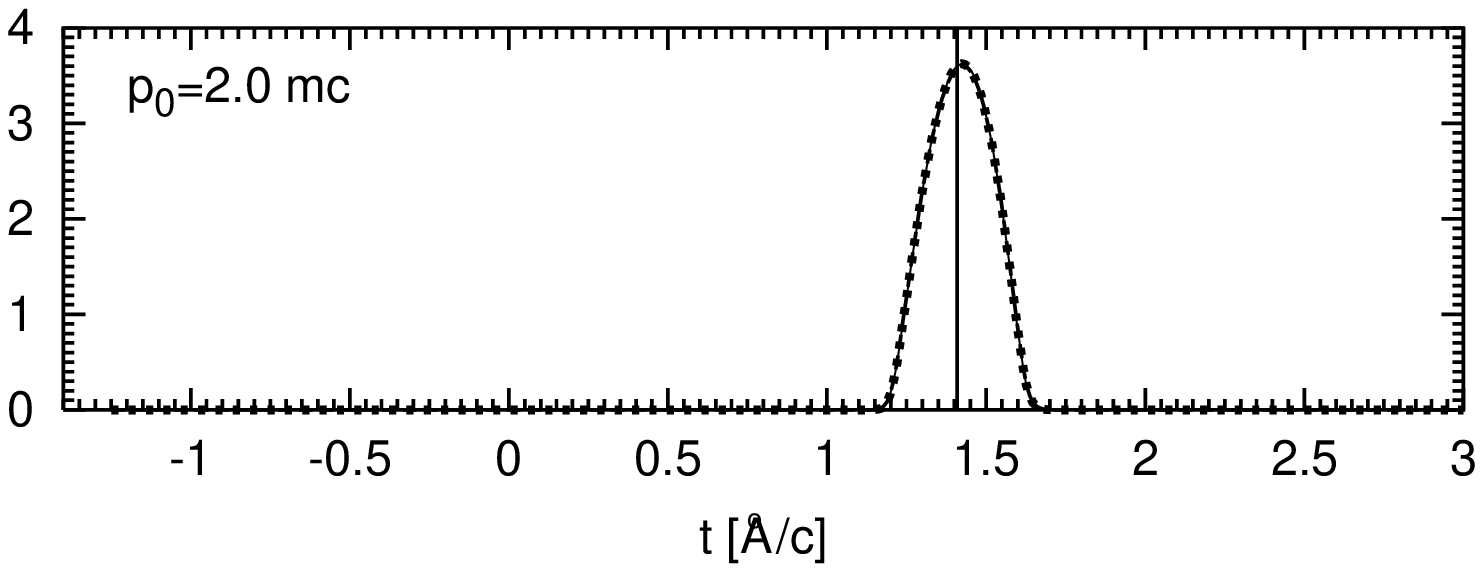}
  \end{flushright}
  \caption{\label{fig_4}Probability densities $\rho_0$ for traversal time in the
      detectors' rest-frame $K_0$, 
      initial state:
      $\Psi_{0,P}$ (small solid line),
      $\Psi_{0,N}$ (big dotted line),
      $\Psi_{0,PN}$ (dashed line),
      particle momentum $p_0$; the
      vertical solid line indicates the traversal time given by classical relativistic
      mechanics}
\end{figure}

\Fref{fig_4} shows the probability densities $\rho_0$ for traversal time in the
detectors' rest-frame $K_0$ with different initial states. The probability
densities have a peak at the classical expected traversal time. Again, we see that
there is nearly no difference if we start with the initial state $\Psi_{0,P}$ or
$\Psi_{0,N}$. There are only small differences with the results
obtained with the initial state $\Psi_{0,PN}$.

Next, we examine the situation in a moving reference frame $K_v$. It moves
with velocity $v$ relative to $K_0$. We calculate the traversal time in the
framework of classical relativistic mechanics:
\begin{eqnarray*}
\tilde{t}_{t,RM} & = & \frac{1}{\fsqrt{1-\frac{v^2}{c^2}}} \cdot \left[
t_{t,RM} - \frac{v}{c^2} (x_2-x_1) \right]
\end{eqnarray*}
We get the same correlation between $t_{t,RM}$ and $\tilde{t}_{t,RM}$ in
classical relativistic mechanics than the correlation between $T_{t,0}$ and $T_{t,v}$ in
our formalism (see \eqref{sec5_Tv_T0}). We get the same result than in
the time-of-arrival case: the good agreement between our results and those
obtained in classical relativistic mechanics exists in all reference frames.

Now, we want to examine how the results depend on
the detector parameters. The particle momentum is fixed at $p_0=0.75
\mc$ and the initial quantum state is $\Psi_{0,P}$.

We start by varying the parameters of the first detector $D_1$. The parameters
of the second detector $D_2$ are fixed at $\Delta x_2 = 0.02 \Ang$ and
$W_2 =1 \times 10^{-3}\mcq$.

\begin{figure}[t]
  \begin{flushright}
    \includegraphics [width=0.46\linewidth]{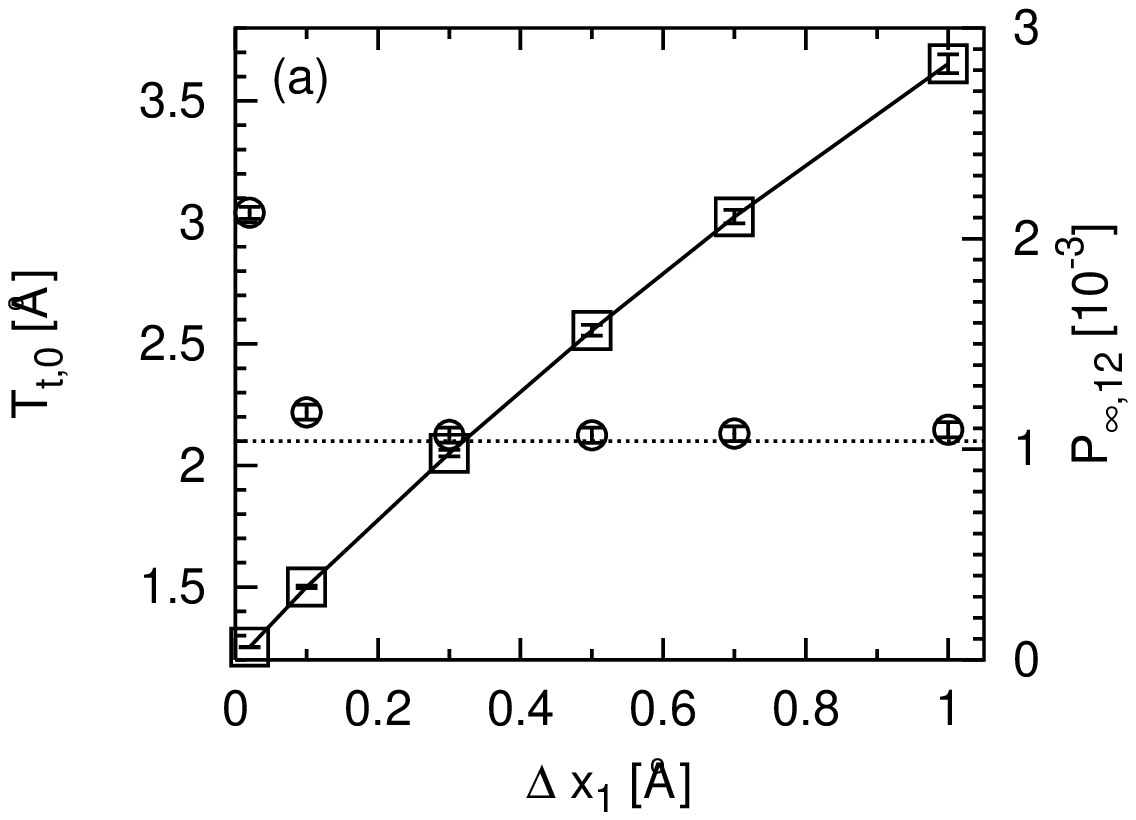}\qquad
    \includegraphics [width=0.46\linewidth]{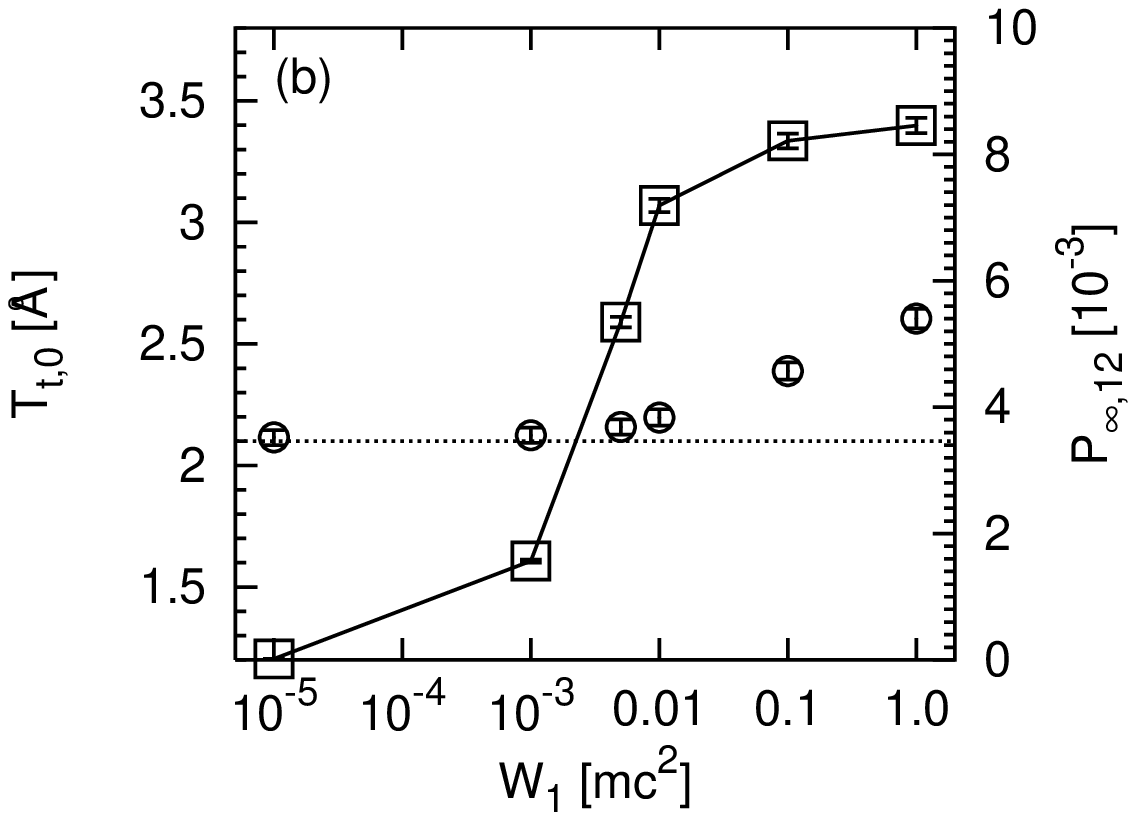}
  \end{flushright}
  \caption{\label{fig_5}Mean traversal time $T_{t,0}$
  (circles with error bars, left axis)
    and probability $P_{\infty,12}$ (boxes with error bars connected
    with a solid line, right axis); initial state $\Psi_{0,P}$ with
    $p_0=0.75mc$; detector $D_2$: $\Delta x_2 = 0.02 \Ang$,
    $W_1=1\times 10^{-3} \mcq$; 
    the dotted line indicates the traversal time deduced from classical relativistic
    mechanics; (a) detector height $W_1 = 1 \times 10^{-3}\mcq$;
    (b) detector width $\Delta x_1 = 0.5\Ang$}
\end{figure}

First, we compute $P_{\infty,12}$ and the expectation value $T_{t,0}$
in $K_0$ for different detector 
widths $\Delta x_1$ while keeping $W_1 =1 \times 10^{-3}\mcq$ fixed
(see \Fref{fig_5}a). We find out that it exists a range of detector
width ($0.3 \Ang \lesssim \Delta x_1 
\lesssim 1.0 \Ang$) for which the expectation value $T_{t,0}$ does not change
in a significant way. But the probability for
two detections $P_{\infty,12}$ increases with increasing detector width
$\Delta x_1$.

In the range $0.3 \Ang \lesssim \Delta x_1 \lesssim 1.0 \Ang$ the forms of the
probability densities $\rho_0$ do not differ in a significant way. The
peaks only become wider with increasing detector width $\Delta x_1$.
If the detector width is very small ($\Delta x_1 = 0.02 \Ang$),
the wave function changes strongly through the detection by $D_1$ and so we get a
qualitatively different probability density $\rho_0$.

Now we fix $\Delta x_1 = 0.5\Ang$ and vary $W_1$ (see
\Fref{fig_5}b). In the case of weakly intrusive detectors $W_1
\lesssim 5 \times 10^{-3} \mcq$, 
the expectation values $T_{t,0}$ do not differ in a significant way. For higher
detectors, the expectation values $T_{t,0}$ increase a bit with increasing
detector height $W_1$. The probability $P_{\infty,12}$ increases with
increasing $W_1$, a fact one expects intuitively.

With increasing detector height $W_1$, the peak of the
probability densities $\rho_0$ is shifted to higher traversal times.

In the last part of this section, we fix the parameters of $D_1$ at $\Delta x_1 =
0.5 \Ang$ and $W_1 = 1 \times 10^{-3} \mcq$ and we vary the parameters $\Delta
x_2$ and $W_2$ of detector $D_2$.
We examine the following pairs of detector parameter 
$\Delta x_2 = 0.02 \Ang$/$W_2 = 1 \times 10^{-3} \mcq$,
$\Delta x_2 = 0.02 \Ang$/$W_2 = 1.0 \mcq$,
$\Delta x_2 = 0.5 \Ang$/$W_2 = 1 \times 10^{-3} \mcq$, and
$\Delta x_2 = 0.5 \Ang$/$W_2 = 1.0 \mcq$.
We find that the resulting probability densities $\rho_0$ and expectation values
$T_{t,0}$ are nearly the same in the first three cases.
The only exception is the case of a very wide and ``height''
detector (last case). In that case, the mean traversal time $T_{t,0}$ is
lower than in the other cases.

The probability $P_{\infty,12}$ grows significantly if one increases the
detector width $\Delta x_2$ or the detector height $W_2$. We get the same
qualitative dependence of $P_{\infty,12}$ on the parameters of detector $D_2$ as
on the parameters of detector $D_1$.

Note that the following fact is true in the case of weakly intrusive detectors
($W_1=W_2=1 \times 10^{-3} \mcq$): the dependence of $T_{t,0}$ on $\Delta x_1$
is ``stronger'' than the dependence on $\Delta x_2$. The reason for this is clear:
changing the width $\Delta x_1$ of the first detector $D_1$ change not only
the first ``detection-time'' but also the form of the wave function after the first
detection.

Summarizing, it exists a wide range of parameters of $D_1$ and $D_2$
for which the mean traversal time does not change
significantly. Remember that the same result was found in the study of
time-of-arrival.


\section{Conclusion}

In \cite{ruschhaupt.2002a}, we have proposed an extension of EEQT describing
one spin $\frac{1}{2}$-particle in a relativistic, covariant way.

In this paper, we have focused on applications of this formalism. 
We have calculated detection times of the particle in
two-dimensional spacetime. The particle
has moved freely except of the influence exerted on it by the detector(s). 

As a first application, we have computed the time at which the particle
arrives at a detector (``free time-of-arrival''). 
We have found out that there exists good
agreement between the expectation values of our simulation and
the results obtained by using classical relativistic mechanics of a free
point particle. Moreover, we have shown that this fact is independent of the
reference frame: the agreement is equally good in all reference frames.
We have considered the situation with different detectors. We have found out that
the good agreement between the results of our algorithm and the results
obtained by using classical relativistic mechanics is not limited to a special value of
detector parameters. It is obtained for a wide range of detector parameters.

As a second application, we have examined the time difference between
two detections of the particle by two detectors, one behind the
other (``free traversal time''). Again, we have obtained good agreement between the simulated
results and those obtained by classical relativistic mechanics in all reference frames.
Moreover, we have shown that there again exist a wide range of detector parameters for
which the mean traversal time does not change significantly.

Summarizing, we have found interesting and meaningful results in these applications of our
relativistic extension of EEQT. So we think that our formalism will be also useful
in future applications, for example in the case, in which the particle
is submitted to the action of a potential barrier.

%

\ack{I would like to thank Ph. Blanchard for many helpful discussions
and the critical reading of the manuscript.}

%

\end{document}